# Multi-degree-of-freedom hybrid optical skyrmions

Jun Yao,[1,*] Yijie Shen,[2,3,*] Jun Hu,[4,†] and Yuanjie Yang[1,‡]

[1]*School of Physics, University of Electronic Science and Technology of China, Chengdu 611731, China.*

[2]*Centre for Disruptive Photonic Technologies, School of Physical and Mathematical Sciences & The Photonics Institute, Nanyang Technological University, Singapore 637378, Singapore.*

[3]*School of Electrical and Electronic Engineering, Nanyang Technological University, Singapore 639798, Singapore.*

[4]*School of Electronic Science and Engineering, University of Electronic Science and Technology of China, Chengdu 611731, China.*

The optical counterparts of skyrmions have recently been constructed with diverse topological types and by different degrees of freedom, such as field, spins, and Stokes vectors, exhibiting extensive potential in modern information science. However, there is currently no method capable of generating multiple types of optical skyrmions in free space. Here, we present a simple approach for realizing hybrid optical skyrmions of electric field vectors, spin angular momentum and Stokes vectors in a same structured light field. We show that a vector beam truncated by an annular aperture can form an electric field skyrmion in the diffracted light field. In the meantime, electric field meron pairs, spin skyrmions and Stokes skyrmions can be generated by tuning spin-orbital coupling of the incident light.

Skyrmions are quasiparticles of a three-component vector field in two dimensions that are topologically protected, characterized by an integer topological invariant known as the skyrmion number [1-4]. Typically, the orientation of the vectors in skyrmions gradually transitions from a central upward (downward) state to an edge downward (upward) state [5,6]. The field configurations of skyrmions cannot be smoothly transformed into another with a different skyrmion number. The concept of skyrmions was initially proposed by Skyrme in the 1960s to describe the stability of hadrons in high-energy physics but it has not been well realized until now [7]. Interestingly, the notion of skyrmions exhibits significant value in condensed matter physics, including quantum Hall systems [8], liquid crystals [9,10], and Bose-Einstein condensates [11]. Particularly in magnetic materials, the compactness and topological stability of the skyrmions make them promising candidates for high-efficiency information storage and transfer [12,13].

The optical counterparts of skyrmions were first realized in surface plasmons by electric field vectors [14]. Subsequently, optical skyrmions have also been formed with photon spins [15-18], magnetic field vectors [19-22], Stokes vectors [23-29], Poynting vectors [30], and pseudospins [31-33]. These quasiparticles, with different vector fields, provide a new degree of freedom for manipulating the light field and have expanded their applications to include sub-wavelength microscopy [34], precision metrology [35], topological Hall devices [31, 36], and quantum information [37]. Optical skyrmions possess tremendous potential in the interaction between light and matter [38-42]. However, up to now, there remains an absence of an approach that can flexibly produce multiple types of optical skyrmions. In particular, the time-varying characteristics of the electromagnetic field pose a great challenge to generating electric field skyrmions [34,43-45].

In this study, we present the generation of multiple optical skyrmion types in free space through annular aperture diffraction. When an annular aperture is illuminated by a vector vortex beam with different spin and orbital quantum numbers, optical skyrmions of electric field vectors, spin vectors, and Stokes vectors can be formed. Despite the time-varying electromagnetic field, the skyrmion number of the electric field skyrmion remains ±1. Our research can inspire further exploration of the interactions between optical skyrmions and matter, as well as facilitate the construction of electron spin skyrmions [46].

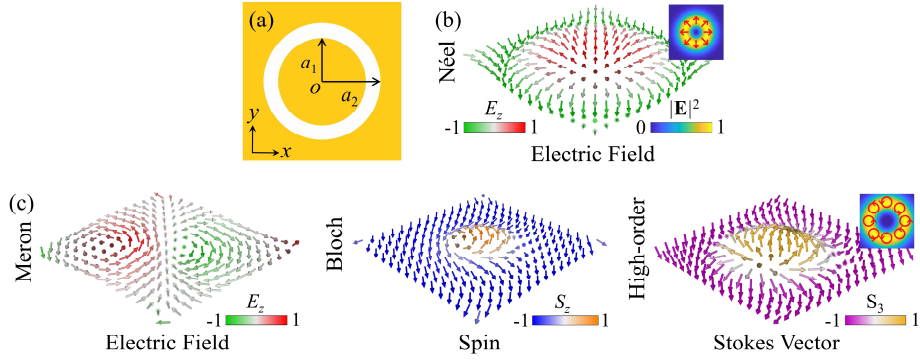

FIG. 1  Optical skyrmions from annular aperture diffraction of vector vortex beams. (a) The annular aperture mask with inner radius $a_1$ and outer radius $a_2$. (b) The Néel-type electric field skyrmion produced from the diffraction of the radially polarized beam ($s = 0$, $l = 0$) shown in the inset. (c) The electric field meron pair, Bloch-type spin skyrmion and high-order Stokes skyrmion produced from the diffraction of the circularly polarized vortex beam ($s = 1$, $l = -2$).

Figure 1(a) shows the annular aperture mask at the plane $z = 0$, illuminated by a vector vortex beam of 532 nm wavelength. The inner and outer radii of the annular aperture are set as $a_1 = 6$ μm and $a_2 = 6.6$ μm, respectively. When the annular aperture is illuminated by the radially polarized beam characterized by the spin quantum number $s = 0$ and the orbital quantum number $l = 0$, a Néel-type electric field skyrmion can be observed in the diffracted light field [Fig. 1(b)]. Interestingly, when the annular aperture is illuminated by the circularly polarized vortex beam with $s = 1$, $l = -2$, the diffracted light field simultaneously displays an electric field meron pair, a Bloch-type spin skyrmion and a high-order Stokes skyrmion [Fig. 1(c)].

Figure 2(a) shows the intensity and electric field vector distributions of the diffracted light field when a radially polarized beam is incident. The diffracted light field is calculated by the finite-difference time-domain method. Due to photonic spin-orbit coupling, the energy within the central region of the diffracted light field is enhanced. The electric field vector distribution within the white rectangle shows adjacent vortex rings with opposite chiralities connected by Bloch points [47], resulting in skyrmion structures in the transverse plane. The in-plane ($xy$) and out-of-plane ($z$) components of the diffracted light field on plane $z = 10$ μm are presented in Fig. 2(b). At position $x = r_0$, the $xy$-component is almost zero. The phase of the $z$-component is shown in Fig. 2(c). The phase difference of $\pi$ between the radial positions $r = 0$ and $r = r_0$ refers to a key characteristic of a skyrmion topological texture. The radius $r_1$ and skyrmion number $N_s$ of the electric field skyrmion over time are presented in Fig. 2(d). Here, the skyrmion number is defined as [48]

$$N_s = \frac{1}{4\pi} \iint_A \mathbf{n} \cdot \left( \frac{\partial \mathbf{n}}{\partial x} \times \frac{\partial \mathbf{n}}{\partial y} \right) dA, \quad (1)$$

where $A$ covers the complete skyrmion and $\mathbf{n} = (n_x, n_y, n_z)$ is a real, normalized, three-component field. Due to an almost zero but nonzero $xy$-component at position $r = r_0$, the radius $r_1$ varies around $r_0$ over time. Inserting a focusing lens behind the annular aperture can maintain a constant radius of the electric field skyrmion (Fig. S1). Near $t = t_1$ and $t = t_2$, the skyrmion number varies between 1 and -1. This phenomenon can be attributed to the fact that, at $t = t_1$ and $t = t_2$, the transient electric fields consist solely of the $xy$- and $z$-component, respectively [Fig. 2(e)]. Figures 2(f$_1$)-(f$_4$) exhibit the dynamic vector configurations of the Néel-type electric field skyrmion over a time cycle. It can be calculated that the skyrmion numbers $N_s = 0.9993, 0.9992, 0.9993,$ and $0.9992$, respectively. $\theta$ is the polar angle of the electric field vectors under local spherical coordinates. The monotonic $\theta$ variation from the center to the periphery in the insets confirms the configuration of skyrmion.

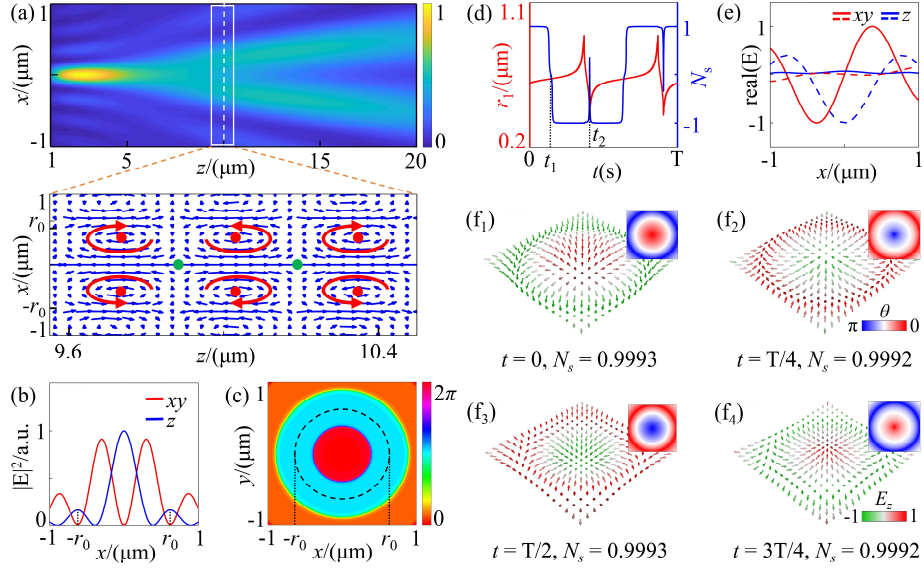

FIG. 2 The electric field skyrmion generated by the annular aperture diffraction of the radially polarized beam. (a) The intensity and electric field vector distributions of the diffracted light field. Red and green dots mark the vortex-type singularities and Bloch points, respectively. (b) Intensity distributions of the $xy$- and $z$-components in the diffracted light field on plane $z = 10$ μm. (c) The phase distribution of the $z$ component. (d) The time-varying radius $r_1$ and skyrmion number $N_s$ of the electric field skyrmion. (e) The transient electric fields at $t = t_1$ (solid lines) and $t = t_2$ (dashed lines). (f$_1$)-(f$_4$) The vector configurations of the Néel-type electric field skyrmion in different phase delays within a time cycle. Insets: the polar angle $\theta$ of the electric field vectors under local spherical coordinates.

Figure 3(a) exhibits the diffracted light field for the circularly polarized vortex beam incident ($s = 1$ and $l = -2$). Spin-orbit coupling induces a non-zero intensity distribution within the central region of the diffracted light field. On the transverse plane at $z = 8.76$ μm, the distribution of electric field vectors exhibits the topological texture of meron pair with $N_s = \pm 0.5$ [Figs. 3(b$_1$)-(b$_4$)]. Notably, the meron pair rotates over time at the wave periods. The distributions of the polar angle $\theta$ in the insets demonstrate the configuration of the meron pair.

Remarkably, the annular aperture diffraction of the circularly polarized vortex beam with $s = 1$, $l = -2$ also simultaneously generates the spin skyrmion and Stokes skyrmion. The $x$-, $y$- and $z$-components of photon spin angular momentum (SAM) along the $x$-axis on the plane $z = 8.76$ μm are shown in Fig. 3(c$_1$). The time-averaged SAM density of the light field is defined as [49]

$$\mathbf{S} = \mathrm{Im}\left(\varepsilon \mathbf{E}^* \times \mathbf{E} + \mu \mathbf{H}^* \times \mathbf{H}\right)/(4\omega), \quad (2)$$

where $\omega$ is the frequency, and $\varepsilon$ and $\mu$ are the permittivity and permeability. Within the region $|x| \leq r_3$, the $x$-component of the SAM almost vanishes with opposite $z$-component directions at positions $x = 0$ and $x = r_3$, indicating a Bloch-type spin skyrmion. The polar angle $\theta$ distribution of the spin vectors is plotted in Fig. 3(c$_2$). From $r = 0$ to $r = r_3$, the direction of spin vectors undergoes a monotonic transition from the upward state to the downward state, which confirms the configuration of a spin skyrmion. Fig. 3(c$_3$) is the vector distribution of the Bloch-type spin skyrmion with $N_s = 0.9979$.

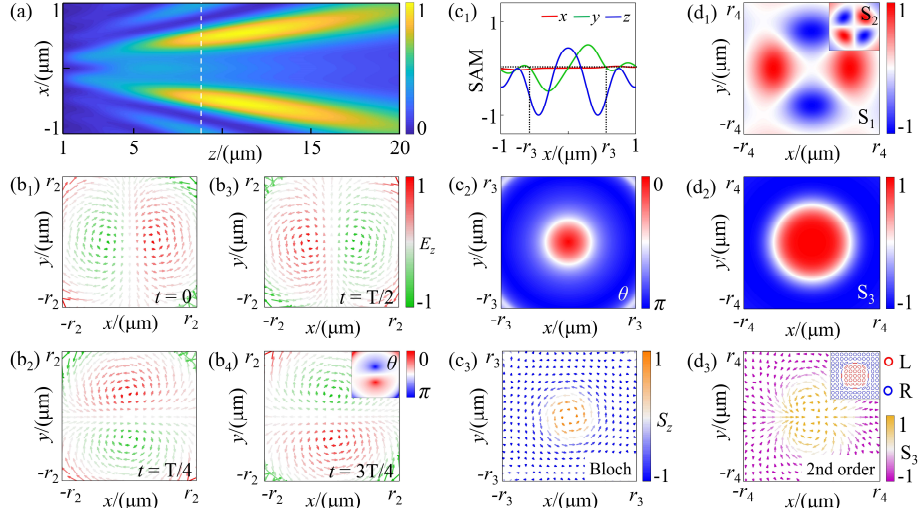

FIG. 3 The multi-degree-of-freedom hybrid skyrmions produced by the annular aperture diffraction of the circularly polarized vortex beam with $s = 1$, $l = -2$. (a) The intensity distribution of the diffracted light field. ($b_1$)-($b_4$) The vector configurations of the electric field meron pair on the plane $z = 8.76$ μm with different phase delays. T is the wave period. Insets: the polar angle $\theta$ of the electric field vectors under local spherical coordinates. ($c_1$) The $x$-, $y$-, and $z$-components of the SAM in the diffracted light field. ($c_2$) The polar angle $\theta$ of the spin vectors. ($c_3$) The Bloch-type spin skyrmion with $N_s = 0.9979$. ($d_1$)-($d_2$) The Stokes parameters $S_1$, $S_2$, and $S_3$ of the diffracted light field. ($d_3$) The Stokes skyrmion with $N_s = 1.9919$. Inset: the polarization state distribution of the diffracted light field. Red and blue represent left-handed and right-handed elliptically polarized states, respectively.

The Stokes skyrmion is solely associated to the xy-component of the light field. The Stokes parameters S1, S2 and S3 are shown in Figs. 3($d_1$)-($d_2$), respectively. It can be concluded that a second-order Stokes skyrmion is constructed. Fig. 3($d_3$) shows the vector configuration of the Stokes skyrmion with $N_s = 1.9919$. The polarization state distribution of the diffracted light field is depicted in the inset of Fig. 3($d_3$). From the center to the edge of the circle region with a radius of $r_4$, the electric field gradually transitions from left-handed circular polarization state to right-handed circular polarization state. Furthermore, by adjusting the observation plane, the azimuthal angle $\varphi$ of the spin vectors can be controlled [Figs. 4($a_1$)-($a_3$)]. Additionally, the Stokes skyrmions exhibit relative rotation [Figs. 4($b_1$)-($b_3$)]. On the planes $z = 5$ μm and 15 μm, rotations of $\phi_1 = -39°$ and $\phi_2 = 40°$ are calculated, respectively, relative to the plane $z = 8.76$μm.

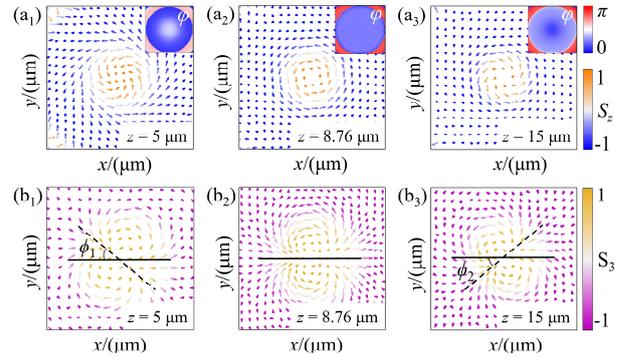

FIG. 4 The spin and Stokes skyrmions on the different observation planes. ($a_1$)-($a_3$) The spin skyrmions on planes $z =$ ($a_1$) 5 μm, ($a_2$) 8.76 μm, and ($a_3$) 15 μm with $N_s = 0.9968$, 0.9979, and 0.9987 respectively. Insets: the azimuthal angle $\varphi$ of the spin vectors under local spherical coordinates. ($b_1$)-($b_3$) The Stokes skyrmions on planes $z =$ ($b_1$) 5 μm, ($b_2$) 8.76 μm, and ($b_3$) 15 μm with $N_s = 1.9854$, 1.9919, and 1.9960 respectively.

The topological configurations of the optical skyrmions in the diffracted light field can be further manipulated by various spin and orbital quantum numbers [29]. The optical skyrmions with $N_s = -1$ can be obtained with $s = -1$ and $l = 2$. In fact, the formation of the optical skyrmions in the diffracted light field is attributed to the diffraction of the light field by the inner edge of the annular aperture. Consequently, the annular aperture is able to be substituted with a circular disk, and both structures are conducive to integration onto the fiber end surface. The annular aperture can also induce spin-orbit coupling in electron beams, suggesting potential application in the construction of electron spin skyrmions. Moreover, the topological magnetic Bloch points can be created by the annular aperture diffraction of the azimuthally polarized beam, which holds significant implications for the investigation of Bloch points in magnetic materials.

In summary, we propose a flexible method for generating multi-degree-of-freedom hybrid optical skyrmions in free space. A Néel-type electric field skyrmion can be achieved through annular aperture diffraction of the radially polarized beam. Furthermore, the electric field meron pair, Bloch-type spin skyrmion and high-order Stokes skyrmion can be simultaneously generated by tuning spin-orbital coupling of the incident light. Our study can facilitate further investigation into the complex light-matter interactions.

*Acknowledgments*—This study is financially supported by the National Natural Science Foundation of China (Nos. 12174047).

*These authors contributed equally to this work.
†hujun@uestc.edu.cn
‡dr.yang2003@uestc.edu.cn